# The DCS Theorem

Greg Slepak
hi@okturtles.com

Anya Petrova
a.petrova.ds@gmail.com

October 4, 2017

**Abstract.** Blockchain design involves many tradeoffs, and much debate has focused on tradeoffs related to scaling parameters such as blocksize. To address some of the confusion around this subject, we present a probability proof of the *DCS Triangle* [1][2]. We use the triangle to show decentralized consensus systems, like blockchains, can have *Decentralization*, *Consensus*, or *Scale*, but not all three properties simultaneously. We then describe two methods for getting around the limitations suggested by the triangle.

## 1 Definitions

A *system* is defined as any set of components (see Decentralization Scope) following *precise rules* in order to provide service(s) to the users of the system. These services constitute the system's *intended behavior*.

In other words, a system $S$ consists of a set of components, called its *scope* $\{S\}$, and a program ("state transition function", $f_S$), that together define the system's *intended behavior*, which means: upon receipt of message $m$, $S$ uses $f_S$ to update the internal state from $s$ to $s'$ and send back reply $y$ within a time interval $S_\tau$.

$$S(t) = \begin{cases} \{S\} & = \{component_1, component_2, \cdots\} \\ f_S(m, s) & = \{s', y\} \end{cases}$$

We note, additionally:

- The *scope* $\{S\}$ may change over time, but there are always several components of a vital type (i.e. "all systems always have at least one *CPU*, one *developer*, and one *user*").
- The system's state $s$ includes all data necessary for the system to compute $f_S$ given a message $m$. The system may limit messages to those that are authorized in some way (in order to prevent denial-of-service).[1]
- $S$ is considered *compromised* if it fails to perform its intended behavior within the interval $S_\tau$.

We will proceed to prove that any single such system may possess, at most, two of three properties:

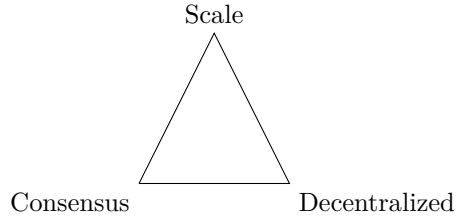

- **Consensus** means the system uses a collective decision-making process ("consensus algorithm") to update the system's state, $s$, which is shared by all *consensus participants*. The result of the consensus algorithm determines the network's accepted output of $f_S$, and whether or not $f_S$ completes within $S_\tau$.
- **Scale** means the system is capable of handling the transactional demands of any competing system providing the same service to the same arbitrary set of users across the globe (*"at scale"*).[2]
- **Decentralized** means the system has no *single point of failure or control* (SPoF). Another way to state this is: if any single element is removed from $\{S\}$, the system continues to perform its intended behavior, and no single component in $\{S\}$ has the power to redefine $f_S$ on its own.

## 1.1 Consensus participants and "full" consensus

The concept of a "consensus participant" is sometimes confused with the concept of a "validator", and in order to understand what the DCS Triangle is saying it's necessary to understand the difference between the two.

Every consensus process has three ingredients: voters (consensus participants), voting rules, and the votes themselves.

In distributed systems, the job of a *validator* is to verify that the voting rules were followed, accepting the outcome of the vote if that is so, and rejecting the outcome otherwise. For example, in the physical world a validator might be responsible for verifying ballot forms were filled out correctly and were cast by registered voters only, but beyond that they do not (generally speaking) have the ability to influence the outcome of the vote.

---

[1] For decentralized systems, this is okay as long as there is no central authority determining who is or isn't authorized.

[2] Examples of "services" include: streaming video, sending messages, maintaining balances on a ledger, etc.

Consensus participants, on the other hand, are the voters themselves, and their job is to not only ensure that voting rules are followed, but to cast a vote on some decision.

In Bitcoin, for example, "miners" are consensus participants whose job is to vote on which transactions are accepted into the blockchain, whereas non-mining "full nodes" are validators only, and their job is to ensure that miners do not produce invalid blocks.

**Definition.** *Consensus participants* are independent entities who each maintain a complete copy of a system's state, and together vote on updates to this shared state.

The notion of a "complete copy of a system's state" is of utmost importance for our proof. In other words, our proof focuses specifically on the strongest notion of "consensus", where each consensus participant has full knowledge of the entire system state, and therefore is able to cast a vote without needing to trust any other participant.

To emphasize this notion of consensus over weaker forms, we'll refer to it as *full* consensus in our theorem.

In *[§3 - Getting around the DCS Triangle](#)*, we'll explore how, by loosening this requirement and treating "consensus" as a spectrum of trust assumptions, it may be possible to design decentralized consensus systems that scale with "good-enough-consensus".

## 1.2 Decentralization *scope* & *relativity*

Implicit to our definition of a *decentralized system* is the idea that the system is not compromised. A non-functioning system does not fulfill its intended behavior, and therefore, by our definition, is not decentralized.

Imagine a decentralized system $S$, whose intended behavior (its purpose) is to maintain the integrity of a database while being responsive to queries. It does so by attempting to eliminate all single points of failure within a given *scope*.

**Definition.** The *scope* of a system refers to all subcomponents and all entities reasonably relevant to a system's functioning.

If we consider the scope of our "decentralized" database to be a computer with two CPUs and two hard disks (one primary, another backup), then we can say $S$ is "decentralized" at $t = 0$ (has no single point of failure). However, if at $t = 1$ one of the hard disk fails, it is no longer decentralized since now there does exist a single component capable of compromising the entire system.

This means:

- Whether or not a system is decentralized can change over time.
- Any system can be called "decentralized" if we define the scope narrowly enough.
- All decentralized systems can be called "centralized" if we define their scope broadly enough.[3]

The narrowing and enlarging of the scope is called the *relativity of decentralization*, and it is why first agreeing on a reasonable definition for a system's scope is vital before deciding whether or not it is "decentralized".

## 1.3 Computational throughput of consensus systems

**Definition.** The *computational throughput* of a consensus system refers to the rate at which the system updates its state by processing all input messages.

We'll use the shorthand $T(S)$ to represent this concept and note three factors that determine its value:

1. The *computational power*[4] of each consensus participant.
2. The amount of time after which the consensus algorithm considers messages to be lost (the *timeout* period).
3. The consensus *threshold* that decides when consensus has been reached (i.e. "how big of a quorum is required").

Note that if the computational power of a consensus participant is significantly less than that of the other participants, they are more likely to be excluded from the deciding quorum for several reasons:

- If there are no network partitions to determine otherwise, fast consensus participants will process messages more quickly and therefore will be first to create a quorum.
- If there are enough fast consensus participants to create a large enough quorum to exceed the system's consensus threshold, then there is no need to wait for the remaining votes of the slow participants.
- Slow consensus participants are more likely than fast consensus participants to hit the system's timeout period for processing and responding to messages, and therefore are more at risk of being excluded from the consensus process entirely.

Therefore, $T(S)$ is a function that is limited by the slowest consensus participants not excluded in the deciding quorum.

---

[3]The entire Internet could be considered centralized if we include the entire solar system as part of the scope. The "single point of failure" could be the Earth itself, its atmosphere, the Sun, etc. Or, perhaps in the not distant future, a single ISP.

[4]This refers to all computational requirements relevant for consensus participation, such as bandwidth, data storage, and processing speed.

## 1.4 Coordination costs

Relevant for our proof is the notion of *coordination costs*, or the difficulty for one entity to engage another and work toward a common goal, because that can result in the formation of a cartel, which in turn violates the requirement that consensus participants be *independent*.

For example, when Bitcoin was first launched, it would be difficult for any miner to find enough collaborating miners to create a cartel with >50% of the hash power, simply because there were many "relevant miners" (consensus participants) distributed all over the world.

Today, however, there are significantly fewer consensus participants in Bitcoin, and it is much easier to (1) identify them, and (2) bring them together to coordinate around some goal. Therefore, we say the coordination costs are lower today than before.

We can approximate the coordination costs $C(S)$ of any consensus system simply as the number of consensus participants:

$$C(S) = \texttt{num\_consensus\_participants}(\{S\})$$

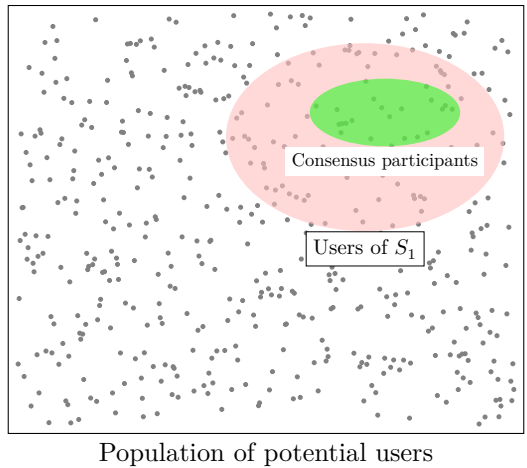

Population of potential users

**Fig. 1:** If $S_1$ is a decentralized consensus system, the DCS Theorem states that as the number of users increases (red circle), the number of consensus participants decreases (green circle).

## 2 Proof

**Theorem 1.** *Decentralized consensus systems centralize at scale when consensus participants maintain full consensus over the entire state of the system.*

We begin with the following axioms accepted as true:

**Axiom 1.** *In any sufficiently large population (at scale), individual access to computational power is distributed unequally. Most have access to average computational power, and a few have access to large amounts.*

Justification: empirically true.

**Axiom 2.** *For any two systems offering the same service to the same large population, the transactional demands of the average user converge at scale.*

Justification: follows from central limit theorem and the law of large numbers.

**Axiom 3.** *Most users of a system do not have the computational power required to store and process all of the messages generated by all of the users of that system at scale.*

Justification: empirically true.[5]

From those axioms, we derive the following lemmas:

**Lemma 1.** *Let $S$ be a decentralized consensus system whose consensus participants maintain full consensus over the system's state. Let $T(S)$ refer to its computational throughput and $c$ refer to the average computational power of all historical consensus participants at any relevant instant in time. At scale, $T(S)$ exceeds $c$, and the more users $S$ obtains, the more $T(S)$ exceeds $c$.*

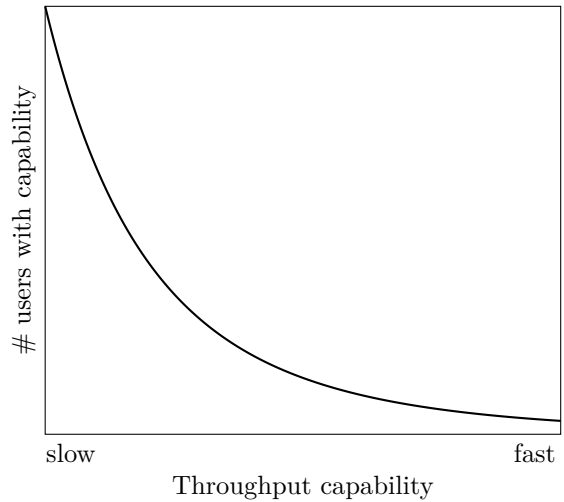

**Fig. 2:** Visualization of (Axiom 1).

*Proof.* This follows directly from Axiom 1, 3, and our definition of a decentralized system, which includes the understanding that for a system to be considered decentralized, it must be uncompromised, and that in turn means it successfully processes all authorized[6] messages from new users within some interval $S_t$. For it to do this, $T(S)$ must exceed $c$, per (Axiom 1) and (Axiom 3). □

---

[5]And perhaps provably true, though such a proof is beyond the scope of this paper.

**Lemma 2.** *Let $S$ be a consensus system as in (Lemma 1). The coordination costs for $S$, $C(S)$, decrease at scale.*

*Proof.* This follows directly from our proof for (Lemma 1) and our definition of $C(S)$. The more $S$ scales, the more $T(s)$ exceeds $c$, and the fewer potential consensus participants are able to participate in consensus. This, in turn, makes it easier for the remaining consensus participants to identify and coordinate with each other. □

**Lemma 3.** *Let $S$ be a consensus system as in (Lemma 1). The probability that $\{S\}$ contains a colluding group capable of censoring transactions increases at scale, and therefore $S$ tends toward centralization at scale.*

*Proof of the Main Theorem.* The final lemma restates our original theorem. As coordination costs decrease (Lemma 2), the probability of a colluding group (a cartel) increases. The presence of a cartel capable of controlling consensus represents a single point of failure *capable* of preventing the system from fulfilling its intended purpose. The definition of a centralized system is one that has a single point of failure. Therefore, we've shown that the probability of the initially decentralized system becoming centralized increases at scale.

It is also worth considering our definition of scale and the implications of (Axiom 2). Per (Axiom 2), when a decentralized consensus system $S_1$ scales to the size of a similar centralized consensus system $S_2$, it will experience the same transactional demands as $S_2$. However, $S_2$ may scale to a size that would guarantee cartel formation in $S_1$ if it were to scale to the same size. Therefore, $S_1$ cannot scale to such a size while remaining decentralized, and therefore $S_1$ cannot satisfy our definition of scale. □

## 3 Getting around the DCS Triangle

As mentioned, the DCS triangle applies to systems employing "full consensus", or in other words, when all consensus participants are required to fully and independently verify the entire state of the system.

It may be possible to "get around" the DCS Triangle by relaxing our definition of consensus. In this section we'll consider two such approaches.

---

[6]See footnote 1 on page 1.

## 3.1 Combining DC and DS systems

Let us suppose we have a DC-system that we wish to scale while preserving its decentralization. An example of such a system is Bitcoin.[3]

Per the triangle, we know that increasing the system's throughput, $T(S)$, via any mechanism that requires all consensus participants to process the additional data, will result in a reduction in the number of independent consensus participants. And so, instead, we may choose to pair our DC-system with a DS-system in some clever way.

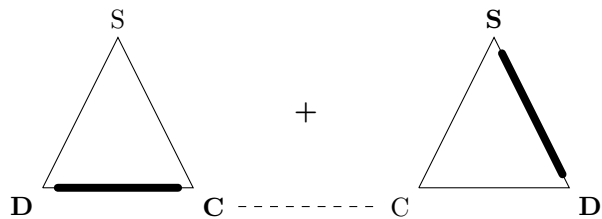

Our DS-system will give us the scale we're looking for, while our DC-system provides a stable and secure source of "ultimate truth" on an as-needed basis. We can connect the two systems in such a way that our DS-system only requires consensus in rare moments, and when it does it may communicate with our DC-system.

The Lightning Network[4] is a real-world example of such a pairing.

## 3.2 Combining multiple DC systems

Yet another possibility is to combine multiple DC systems to create a super-system of DC *groups.*

This approach explores a middle-ground within the DCS triangle, and is the approach taken by systems like OmniLedger.[5]

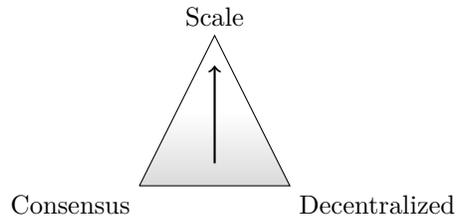

Also known as *sharding,* each group (or *shard*) of consensus participants no longer has complete knowledge of the entire system state, and therefore must (at least partially) trust the other consensus groups. Transparency techniques, such as merkle tree logs, make it possible to minimize the amount of "faith" groups must place in each other.

Overall system consensus is progressively "sacrificed" as the system scales, but only in small, manageable increments. If the system does not need much inter-group consensus, it can scale significantly without issue. If necessary, a DS-system can be added for additional scale.

# 4 Acknowledgements

Thanks to Trent McConaghy and Andrea Devers for their feedback.

# References

[1] T. McConaghy, "The DCS Triangle," 10-Jul-2016. https://medium.com/the-bigchaindb-blog/the-dcs-triangle-5ce0e9e0f1dc.

[2] G. Slepak, "Slepak's Triangle," *Rebooting Web-of-Trust*, 17-Oct-2016. https://github.com/WebOfTrustInfo/rebooting-the-web-of-trust-fall2016/blob/master/topics-and-advance-readings/Slepaks-Triangle.pdf.

[3] S. Nakamoto, "Bitcoin: A Peer-to-peer Electronic Cash System." Oct-2008. https://bitcoin.org/bitcoin.pdf.

[4] J. Poon and T. Dryja, "The Bitcoin Lightning Network: Scalable Off-Chain Instant Payments," 14-Jan-2016. https://lightning.network/lightning-network-paper.pdf.

[5] E. Kokoris-Kogias, P. Jovanovic, L. Gasser, N. Gailly, E. Syta, and B. Ford, "OmniLedger: A Secure, Scale-Out, Decentralized Ledger via Sharding," 2017. https://eprint.iacr.org/2017/406.pdf.